\documentclass[preprint]{emulateapj}

\def\hi{H{\textsc i} }
\def\nhi{$N$(H\textsc{i})}
\def\co{$^{12}$CO}
\def\co13{$^{13}$CO}
\def\co18{C$^{18}$O}
\def\cm3{cm$^{-3}$}
\def\cm2{cm$^{-2}$}

\def\i{\textit}

\def\cm2{cm$^{-2}$}

\def\kms{km s$^{-1}$}

\def\nh3{NH$_3$}
\def\n2h{N$_2$H$^+$}
\def\hc3n{HC$_3$N}
\def\h2{H$_2$}
\def\nh{n(H$_2$)}
\def\cp{C$^+$}

 \def\pasa{PASA}

\usepackage{color}
\usepackage{cancel,soul,ulem}

\slugcomment{ }
\usepackage[breaklinks,colorlinks,urlcolor=blue,citecolor=blue,linkcolor=blue]{hyperref}

\usepackage{enumerate}   
 \usepackage[flushleft]{threeparttable}  
 
\shorttitle{OH Survey along Sightlines of  Galactic Observations of Terahertz C+  }
\shortauthors{Tang et al.}

\begin{document}


\title{OH Survey along Sightlines of  Galactic Observations of Terahertz C+  }

\author{Ningyu Tang\altaffilmark{1,2}, 
Di Li\altaffilmark{1,3}, 
Carl Heiles\altaffilmark{4},
Nannan Yue\altaffilmark{1,2}, 
J. R. Dawson\altaffilmark{5, 6},
Paul F. Goldsmith\altaffilmark{7},
Marko Kr\v{c}o\altaffilmark{1},
N. M. McClure-Griffiths\altaffilmark{8},
Shen Wang\altaffilmark{1,2},
Pei Zuo\altaffilmark{1,2},
Jorge L. Pineda\altaffilmark{8}
and 
Jun-Jie~Wang \altaffilmark{1}
} 


\altaffiltext{1}{National Astronomical Observatories, CAS, Beijing 100012, China; Email: nytang@nao.cas.cn, dili@nao.cas.cn}
\altaffiltext{2}{University of Chinese Academy of Sciences, Beijing 100049, China}
\altaffiltext{3}{Key Laboratory of Radio Astronomy, Chinese Academy of Science}
\altaffiltext{4}{Department of Astronomy, University of California, Berkeley, 601 Campbell Hall 3411, Berkeley, CA 94720-3411}
\altaffiltext{5}{Department of Physics and Astronomy and MQ Research Centre in Astronomy, Astrophysics and Astrophotonics, Macquarie 
 University, NSW 2109, Australia}
\altaffiltext{6}{Australia Telescope National Facility, CSIRO Astronomy and Space Science, PO Box 76, Epping, NSW 1710, Australia}
\altaffiltext{7}{Jet Propulsion Laboratory, California Institute of Technology, 4800 Oak Grove Drive, Pasadena, CA 91109, USA}
\altaffiltext{8}{Research School for Astronomy \& Astrophysics, Australian National University, Canberra, ACT 2611, Australia}

\begin{abstract}
We have obtained OH  spectra of four transitions in the $^2\Pi_{3/2}$ ground state, at 1612, 1665, 1667, and 1720 MHz, toward  51 sightlines that were observed in the  Herschel project Galactic Observations of Terahertz C+.  The observations cover the  longitude range of (32$^\circ$, 64$^\circ$) and (189$^\circ$, 207$^\circ$) in the northern Galactic plane.  All of the diffuse OH  emissions  conform to the so-called `Sum Rule' of the four brightness temperatures, indicating optically thin emission condition for  OH  from diffuse clouds in the Galactic plane. The column densities of  the H{\sc i} `halos'  \nhi\  surrounding molecular clouds  increase monotonically with OH column density, $N\rm(OH)$, until saturating  when $N(\rm\hi)=1.0\times 10^{21}$ \cm2 and $N\rm (OH) \geq 4.5\times 10^{15}$ \cm2, indicating the presence of molecular gas that cannot be traced by H{\sc i}. Such a  linear correlation, albeit weak, is suggestive of \hi halos'  contribution to the UV shielding required for molecular formation. About 18\%  of OH clouds have no associated CO emission (CO-dark) at a  sensitivity of 0.07 K but are associated with \cp emission.  A weak correlation exists  between \cp\ intensity and OH column density for CO-dark molecular clouds. These results imply  that  OH seems to be a better tracer of molecular gas than CO in diffuse molecular regions.  

\end{abstract}



\section{Introduction}
\label{sec:introduction}

The hydroxyl radical (OH) is a relatively abundant, simple hydride, and thus a potentially important  probe of interstellar medium (ISM) structure. It was first detected in absorption against continuum sources \citep{weinreb63} and then in emission toward interstellar dust clouds  (Heiles 1968).  A large number of studies have revealed the widespread existence of OH throughout dense, dusty clouds  \citep{turner71,turner73,crutcher77}, high-latitude translucent clouds  \citep{grossmann90,barriault10,cotten12}, and diffuse regions outside the CO-bright molecular clouds  \citep{wannier93,allen12}.

Four 18 cm ground-state transitions of OH at 1612, 1665, 1667, and 1720 MHz can be readily observed in L band.  Local thermodynamic equilibrium (LTE) was initially considered to be valid for these four transitions \citep[e.g.,][]{heiles69}.  Under optically thin assumption, LTE   implies the ratios $T_{\rm A}(1612):T_{\rm A}(1665):T_{\rm A}(1667):T_{\rm A}(1720)=1:5:9:1$.  Subsequent observations revealed anomalies in satellite (1612, 1720 MHz) and main (1665, 1667 MHz) lines  \citep[e.g.,][]{turner73,guibert78,crutcher79}.  On-source absorption and off-source emission observations toward continuum sources have been used to obtain the optical depth and excitation temperature of each transition independently. The aforementioned surveys have found non-LTE gas with a typical excitation temperature difference of $|\Delta T  \rm_{ex}|\sim 1-2 $ K for the OH main lines \citep[e.g.,][]{rieu76,crutcher77,crutcher79,dickey81}.

Recent observations have shown that CO, the widely used tracer of \h2, does not trace  molecular gas well  in regions with intermediate extinctions $0.37-2.5$ mag \citep[e.g.,][]{planck11}. We refer to such regions as diffuse, or translucent clouds, and when appropriate, $\rm H\textsc{i}\textnormal{-}H_2$ transition regions, throughout this work.  
OH and C$^+$ are key initiators of the chemistry that leads to CO in  diffuse and translucent regions through the reactions \citep{van88} 

$\rm C^+ +OH \rightarrow CO^++H$,

$\rm CO^++H_2 \rightarrow HCO^+ +H$,
 
$\rm HCO^++e \rightarrow CO+H$.  
 
 OH has been detected toward the outer shells, also referred to as `halos',  of  molecular clouds with low CO abundances \citep[e.g.,][]{wannier93,allen12}, and clouds toward continuum sources \citep{li15}.

In order to improve the understanding of the distribution of OH and of the ISM traced by OH, large and sensitive  surveys of  OH  in diffuse gas are necessary.   The first ``blind" survey of diffuse OH taken by \citet{penzias64} was unsuccessful. \citet{turner79} carried out OH survey near Galactic plane with a  sensitivity of 0.18 K. The surveys with high sensitivity (a few mK RMS) by
\citet{allen12,allen15}  covered  regions  ($l,b$)=(108.0$^\circ$, 5.0$^\circ$) and (105.0$^\circ$, 1.0$^\circ$) with the 25 m radio telescope of the Onsala Space Observatory and Green Bank telescope, respectively. The Southern Parkes Large-Area Survey in Hydroxyl (SPLASH)  covered ($l,b$)=(334$^\circ$-344$^\circ$, -2$^\circ$-2$^\circ$) in a pilot region, and will cover $l$ ranges of (332$^\circ$, 10$^\circ$), $|b|\le 2^\circ$ including some additional coverage of higher altitude around the Galactic centre (Dawson, in prep).  In these regions,  observations of  C$^+$  are not available. 

In this work, we adopted a more limited and focused approach by following up on Galactic Observations of Terahertz C+ (GOTC+) survey  \citep{langer10, pineda13} with OH observations.  With a data set of the three important tracers of molecular gas, C$^+$, OH, and CO, we here examine their correlation and their relative efficiency in tracing molecular gas. 

This paper is organized as follows: In sections \ref{sec:observations} and  \ref{sec:reduction_detection}, we describe the observations and data reduction of OH and associated spectral data. In section \ref{sec:analysis} we show procedures for gaussian decomposition and analysis of OH, $\rm \hi$, and CO column density. The results are presented in section \ref{sec:results}. In section \ref{sec:discussion} we provide a discussion of OH column density and atomic/molecular transition. In section \ref{sec:conclusion} we provide the conclusions from our study.

\section{Observations and Data}
\label{sec:observations}
\subsection{OH Observations}
\label{sec:OH_obs}
There are 92 sightlines of the GOTC+ project that are covered  by the Arecibo telescope. We chose observed  sightlines based on the following three criteria:  (1) if the sightline can be observed  for at least half an hour,  (2) if there exist  ``CO-dark''  candidates toward the sightline, and (3) if there exist abundant \hi self-absorption features, C$^+$, and CO emission toward the sightline.  Criteria (1) and (2) have higher  priority.  All sightlines satisfying criteria (1) and (2) have been observed.

In this  survey, OH spectra toward 51 sightlines were obtained. The positions  of these sightlines covering 43 points in the Galactic longitude range of (32$^\circ$, 64$^\circ$) (range A) and 8 points in the Galactic longitude range of (189$^\circ$, 207$^\circ$) (range B) are shown in Figure\ \ref{fig:oh_count_plot}.

The OH observations were carried out with the Arecibo telescope in two periods,
September 15th to November 7th, 2014 and February 26th to  March 3rd, 2015. The observations were made with the
Interim Correlator backend with bandwidth of 3.125 MHz, providing a
velocity resolution of 0.28 km s$^{-1}$ at 1.66 GHz. The integration
time for each sightline was half an hour. To reduce the effect of radio
frequency interference (RFI) and the instability of the receiver and to
avoid difficulty in choosing a clean ``OFF'' position, we developed an observation script
that changes the central reference velocity by 200 km
s$^{-1}$ every 15 min.  This is equivalent to frequency switching, which
is not supported in Arecibo.

\subsection{CO Observations}
\label{sec:CO_obs}

Corresponding $^{12}$CO(1-0) and $^{13}$CO(1-0) observations were made with the Delingha 13.7m telescope between May 4th and 10th, 2016. The 13.7m telescope, located in northwestern China, has an angular resolution of 1 arcmin (FWHP, Full Width at Half Power) at 115 GHz. The  system temperature varied from 250 K to 360 K with a typical value of 300 K during the observations.  The observations were taken using position switching. The total observation time per target was 30 min or 45 min depending on the system temperature. The backend  has a 1 GHz bandwidth and 61 kHz spectral resolution, corresponding to a velocity resolution of 0.16 km/s at 115.271 GHz. 

\subsection{Archival  C$^+$ and  \textrm{H{\sc i}} Data }
\label{sec:archive_data}
C$^+$ data were obtained from the GOT C+ project  \citep{pineda13,langer14}. The data have already been smoothed into a channel width of 0.8 km s$^{-1}$ with an average rms noise of 0.1 K. 

The H{\sc i} data representing brightness temperature were taken from the Galactic Arecibo L-band Feed Array H{\sc i}   \citep[GALFA-H{\sc i};] []{peek11} with a  noise level of 0.33 K in a 0.18 km s$^{-1}$ channel. 

\section{Data Reduction and Processing}
\label{sec:reduction_detection}

\subsection{Description of Data Reduction and Processing}
\label{sec:data_reduction}

The OH data were reduced with our IDL procedures. 
Scans with obvious RFI were firstly removed by checking the correlation map of the data. 
RFI was further checked by comparing averaged spectra in two separate 15 min
observations. This is especially important for the 1612 MHz spectra,
which are significantly affected by RFI. After deriving the bandpass
spectrum, we ignored  the edges of the spectrum where gain of the bandpass
varies and only fitted middle part of the spectrum. Spectral channels
with obvious OH lines were marked to avoid being included in the
bandpass fit. Most of the
bandpass spectra are flat and can be fitted with a first-order
 polynomial. The other spectra were fitted with higher-order
polynomials. Weak OH emission/absorption lines with wide velocity widths
(full width at half maximum $> 8$ km s$^{-1}$) may be  missed during
this step. The final noise level is 35 mK in a 0.28 km s$^{-1}$ channel.

A main beam efficiency of 0.52 was used to transform CO antenna temperatures to main-beam brightness temperatures.  The {\tt GILDAS}\footnote{{\tt http://www.iram.fr/IRAMFR/GILDAS}} software was used for baseline fitting and spectral smooth of CO data. The CO spectra were smoothed  to 0.32 km/s to reach a  velocity resolution comparable to that of OH data. The final noise level of main-beam brightness temperature is $\sim 70$ mK for $^{12}$CO(1-0) in  0.32 km s$^{-1}$ and $\sim 40$ mK for $^{13}$CO(1-0) in  0.33 km s$^{-1}$ channel width. 

The H{\sc i} data were smoothed to a velocity
resolution of 0.36 km s$^{-1}$ that is comparable to the velocity
resolution of OH and CO data. The rms noise level after smoothing is 0.23 K.

\subsection{Detection Statistics}
\label{sec:detection}

\begin{figure}
\centering
  \includegraphics[width=0.47\textwidth]{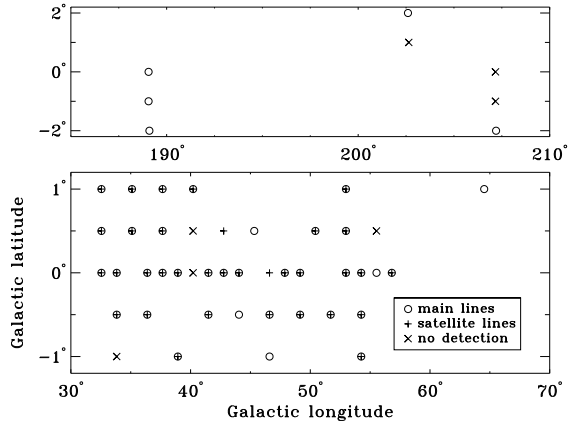}
  \caption{ Detection information toward observed 51 sightlines in different survey region. The top and bottom panel show sightlines in the Galactic longitude range of (185$^\circ$, 210$^\circ$) and (30$^\circ$, 65$^\circ$), respectively. Circle, plus, and cross signs represent  detection of OH main lines, detection of OH satellite lines, and no detection, respectively.}
\label{fig:oh_count_plot}
\end{figure}

With a rms of $\sim 35$ mK,  the detection statistics of the 4 OH lines are displayed in Figure \ref{fig:oh_count_plot}. OH emission/absorption is detected in 44 of 51 sightlines. OH main lines appear in 9 of 44 sightlines  alone while OH satellite lines appear in  2 of 44 sightlines alone. 

The detection rate of OH main and satellite lines varies depending on their locations in
the Galaxy. No OH satellite lines were  detected in the outer Galaxy.  Figure \ref{fig:oh_count_plot}  indicates that the
detection rate of OH lines (including both the main and satellite lines) in
the outer galaxy is 62.5\%, much smaller than that of 93.0\% in the
inner galaxy. This is consistent with the fact that  the
amount of CO-bright molecular gas in the outer galactic plane is smaller than that in the
inner galactic plane  \citep[e.g.,][]{dame01}.  Absorption features are commonly present in OH main
lines in the inner Galaxy  even though there is no H II region in the
beam. But absorption features are absent in OH main lines in the outer Galaxy, indicating a lower level of continuum background in
the  outer galaxy.

\section{Analysis}
\label{sec:analysis}

\subsection{Gaussian Decomposition }
\label{sec:decomp}

We developed  an IDL script to decompose OH, \cp and CO spectra. 
This script uses the classical nonlinear least squares technique,
which utilizes analytically-calculated derivatives, to iteratively
solve for the least squares coefficients. For each spectrum, the number of Gaussian components
was fixed. Initial guesses of each Gaussian
component were required.  The decomposition results were then checked by eye. 

\begin{figure}
\centering
  \includegraphics[width=0.47\textwidth]{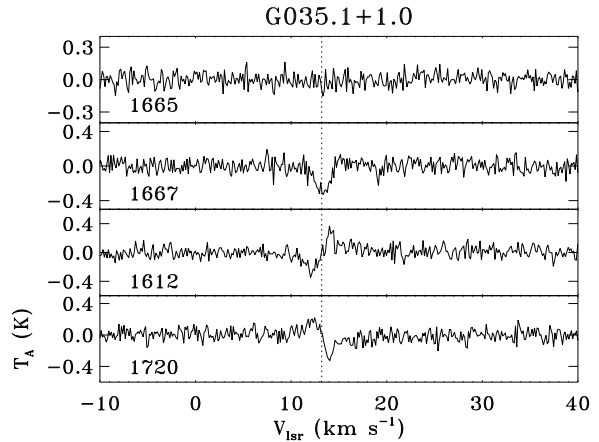}
  \caption{ An example of OH spectra that show clear flip of satellite lines. The  dotted line represents the fitted gaussian central velocity of 1667 MHz, 13.2 km s$^{-1}$. This corresponds to  the velocity where the 1612 and 1720 MHz flips occur. }
\label{fig:sate_anomaly}
\end{figure}

We fit the OH profiles first.  In general, central velocities of the four
OH  lines should be the same for a cloud. A switch from
emission to absorption as a function of velocity in OH satellite lines exists in some clouds. In
these clouds, the central velocities of the main lines  are the same as the
cross  points of the satellite lines. An example is shown in Figure
\ref{fig:sate_anomaly} and discussed in Section
\ref{sec:oh_column_density_dis}.  This always occurs in the clouds near H
II  regions  and can be explained by infrared pumping of the $^2\Pi_{3/2}
J=5/2$ level \citep{turner73,crutcher77}. We treat this kind of
feature as a single component.  

The central velocities of derived OH
components were adopted as the initial guess for decompositions of C$^+$ and CO data.

\begin{figure*}
\begin{center}
  \includegraphics[width=0.8\textwidth]{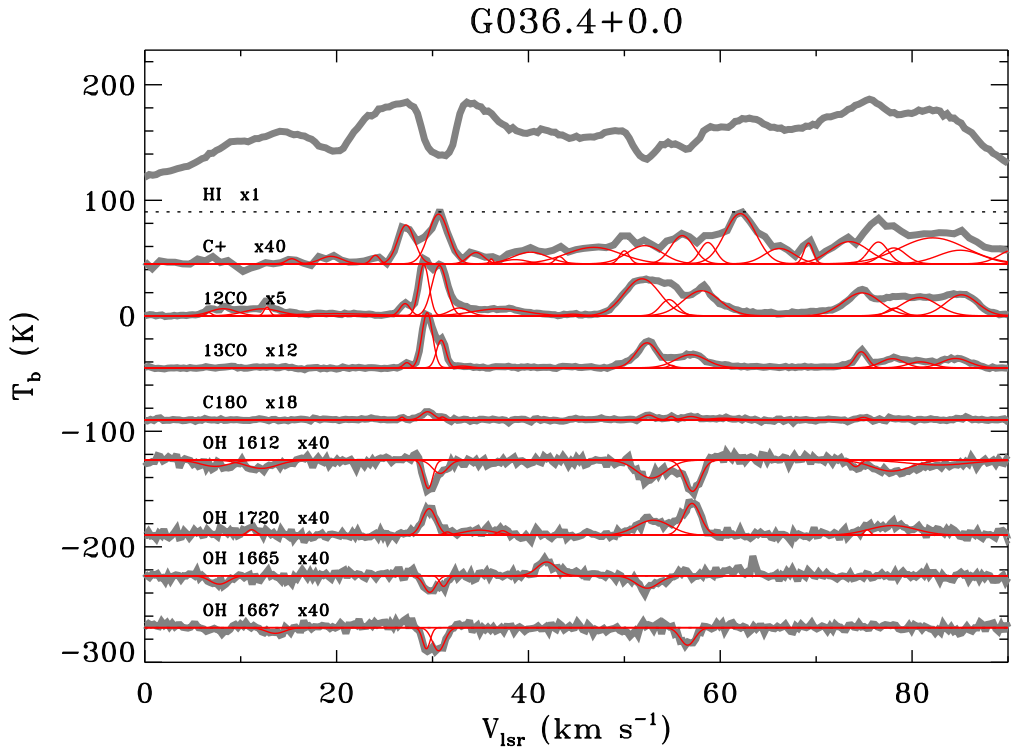}
  \caption{ \hi, C$^+$, CO, and OH spectra toward G036.4+0.0. Red solid lines are the gaussian fits to individual velocity component. }
\label{fig:g036.4+0.0_spec}
\end{center}
\end{figure*}

Finally, 151 cloud components with OH emission or absorption lines
were identified. An example is shown in Figure \ref{fig:g036.4+0.0_spec}.

\subsection{OH Column Density}
\label{sec:oh_col_den}

The brightness temperature ratio between the 4 OH lines
(T$\rm_A^{1612}$:T$\rm_A^{1665}$:T$\rm_A^{1667}$:T$\rm_A^{1720}$) is
1:5:9:1 under assumptions of LTE and optically thin emission \citep[e.g.,][]{robinson67}.
  An anomalous ratio of OH lines that  deviates from the
1:5:9:1 ratio cannot be explained by optical depth effects. An OH anomaly implies
non-LTE conditions  leading to differential excitation  of 4
OH lines. Satellite line anomaly is seen more often than main line
anomaly. The main line transitions occur
between levels with the same total angular momentum quantum number (F). For
satellite lines, transitions occur between energy levels with different
F, which are easily affected by non-thermal excitation \citep{crutcher77}.
Inversion of satellite lines is commonly seen without inversion of main
lines  (see Figures \ref{fig:sate_anomaly}  and 
\ref{fig:g036.4+0.0_spec} for examples), making it
difficult to calculate the OH column density with satellite lines. We
thus calculated OH column densities only for clouds with main line
emission.

The radiative transfer  of the main lines in LTE can be written as
\begin{equation}
T_{\rm mb}^{1665}=F_{\rm b}(T_{\rm ex}-T_{\rm bg})(1-e^{-\tau_{1667}/1.8}),  
\label{eq:65ta}
\end{equation}
\begin{equation}
T_{\rm mb}^{1667}=F_{\rm b}(T_{\rm ex}-T_{\rm bg})(1-e^{-\tau_{1667}}),
\label{eq:67ta}
\end{equation}
where $T_{\rm mb}^{1665}$ and $T_{\rm mb}^{1667}$ are the brightness temperatures of 1665 and 1667 MHz lines, respectively, $F\rm_b$ is the beam filling factor, $T_{\rm ex}$ is the excitation temperature, and $T_{\rm bg}$ is the background continuum temperature at 1.6-1.7 GHz. In high latitude regions, $T\rm_{bg} \sim 3.1$ K at 1.6 GHz, which is the sum of cosmic microwave background (CMB) of 2.73 K \citep{mather94} and Galactic synchrotron emission of $\sim 0.4$ K extrapolating from 408 MHz survey \citep{haslam82}  with a spectral index of 2.7 \citep{giardino02}. 

The contribution from continuum sources (e.g., H II regions) becomes
important at low latitudes, especially the Galactic plane in this
survey.  The HIPASS 1.4 GHz  continuum survey \citep{calabretta14}
was used to estimate continuum emission at 1.6-1.7 GHz toward each
sightline.  We subtracted 3.3 K (the sum of 2.73 K CMB and $\sim 0.6$ K
from Galactic synchrotron emission at 1.4 GHz; e.g.,\ \citet{reich86}) from
HIPASS data and then estimated values at 1612, 1666, and 1720 MHz with
a spectral index of $\sim 2.1$ found in the SPLASH survey. A fraction
factor $p\rm_c$ ($0<p\rm_c< 1$) was utilized to derive continuum
contribution behind OH cloud. With the assumption that the continuum
contribution is uniformly distributed along the sightline across the
Milky Way, $p\rm_c$ is represented as $ (d_{\rm sightline}-d_{\rm
  cloud})/d_{\rm sightline}$, in which $ d_{\rm cloud}$ is distance to
OH cloud and $d_{\rm sightline}$ is the sightline length across the Milky
Way. During the calculations, we applied the Milky Way rotation curve in \citet{brand93}
 and a maximum galactocentric radius of 16 kpc. The
values of $p\rm_c$ vary from 0.48 to 0.98 with a median of 0.90.
Finally, a correction of 3.1 K was added back to derive $T\rm_{bg}$ at
1.6-1.7 GHz. The uncertainties are discussed in the end of this section.  

The OH column density $N(\rm OH)$ can be calculated with the following two general equations   \citep{turner71,liszt96}  
\begin{equation}
  N({\rm OH})=4.07\times 10^{14} cm^{-2} \frac{T_{\rm ex}}{T_{\rm ex}-T_{\rm bg}}  f_{\tau} f_{\rm ex} \int T_{\rm mb}(1665)d\upsilon,  
\label{eq:65_col}
\end{equation}
\begin{equation}
  N({\rm OH})=2.26\times 10^{14} cm^{-2} \frac{T_{\rm ex}}{T_{\rm ex}-T_{\rm bg}} f_{\tau} f_{\rm ex} \int T_{\rm mb}(1667)d\upsilon,   
\label{eq:67_col}
\end{equation}
where $T_{\rm mb}(1665)$ and $ T_{\rm mb}(1667)$  are the main beam brightness temperatures of the 1665 MHz and 1667 MHz lines, respectively. $f_{\tau}=\int \tau d\upsilon/\int(1-e^{-\tau})d\upsilon$ is the correction factor for the optical depth $\tau$ of the OH transitions. The correction factor for $T_{\rm ex}$, $f_{\rm ex}=(h\nu/kT_{\rm ex})/(1-e^{-h\nu/kT_{\rm ex}})$, approaches 1 when $T\rm_{ex} \gg 0.08$ K. 

In LTE, the ratio between brightness temperature of main lines ($R_{1667/1665}=T_{\rm A}^{1667}/T_{\rm A}^{1665}$) varies between 1.8 for optically thin conditions and 1.0 for infinite optical depth  \citep{heiles69}. When $R_{1667/1665}$ was in the range of [1.0, 1.8], the combination of  equation\ \ref{eq:65ta} and equation\ \ref{eq:67ta} can solve for $T\rm_{ex}$ and $\tau_{1667}$ simultaneously. Then $T\rm_{ex}$ and $\tau_{1667}$ can be  inserted into equation \ref{eq:65_col} or \ref{eq:67_col} to solve for N(OH). Previous OH observations have revealed ubiquitous  anomalies between excitation temperatures of the main lines. Non-LTE excitation can lead to  ratios  mimicking LTE range \citep{crutcher79}. Beside this, LTE calculations are limited by satisfaction of sum rule, which implies small optical depth as described in Section \ref{sec:sumrule}.

\begin{figure}
\centering
  \includegraphics[width=0.47\textwidth]{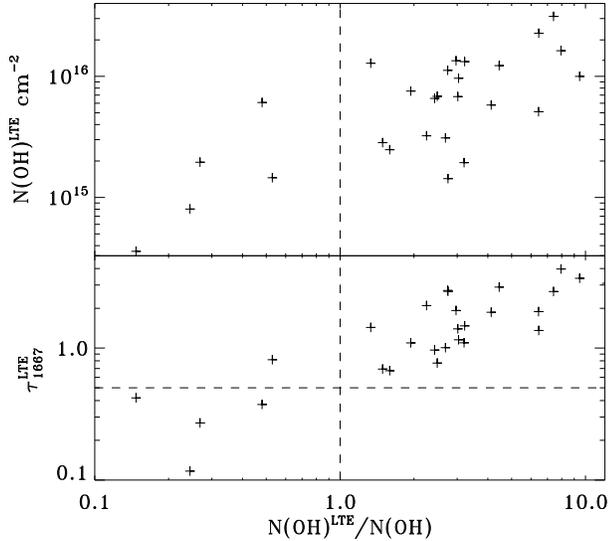}
  \caption{ Comparison between LTE and non-LTE
      calculations for 29 clouds having LTE line ratios.  $N$(OH)$\rm^{LTE}$
      and $\tau\rm_{1667}^{LTE}$ represent total OH column density and
      optical depth of 1667 line using the LTE assumption,
      respectively. N(OH) represents total OH column density using
      non-LTE assumption.  The vertical dashed line represents
      N(OH)$\rm^{LTE}$/N(OH) ratio of 1. The horizontal dashed line
      represents $\tau\rm_{1667}^{LTE}=0.5$. }
\label{fig:lte_nonlte}
\end{figure}

The values of $R_{1667/1665}$ in 29 OH clouds are in the LTE range. As shown
in Figure \ref{fig:lte_nonlte}, the values of $\tau_{1667}$ in 4
clouds are smaller than 0.5 with the LTE assumption. With
consideration of satisfying sum rule implying optically thin as
described in Section \ref{sec:sumrule}, we adopted LTE calculation
results for these 4 clouds. The method for non-LTE OH clouds in case 3
described below was adopted to calculate OH column densities of the  remaining
25 clouds.  As shown in Figure \ref{fig:lte_nonlte}, LTE assumption
generally leads to higher optical depth, $\tau\rm_{1667}^{LTE}> 1.0$ and
larger OH column density, $N$(OH)$\rm^{LTE}>1.0\times 10^{16}$ cm$^{-2}$
than non-LTE assumption.

\begin{figure}
  \includegraphics[width=0.47\textwidth]{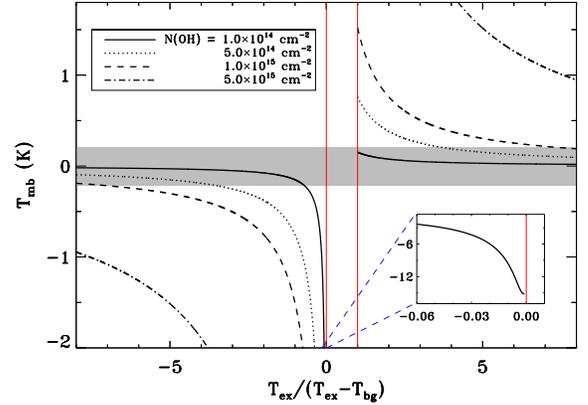}
  \caption{Brightness temperature of the 1665 MHz line as a function of the OH column density, N(OH), and the temperature factor $T\rm_{ex}$/($T\rm_{ex}$-$T\rm_{bg}$). Positive excitation temperature $T\rm_{ex}$ was considered, leading to temperature factor range of (-$\infty$,0) and (1,$\infty$). The two red vertical lines represent values of 0 and 1 for the temperature factor. A typical background continuum temperature of 8 K and FWHM of 1.5 km s$^{-1}$ in this  survey were adopted, and optically thin condition was assumed. The grey shaded region covers the parameter space within the 3$\sigma$ detection limit of 0.21 K. OH lines   within that region cannot be detected in our survey.  When we zoom to  the temperature factor in the range of [-0.06, 0.01] for $N(\rm OH)=1.0\times 10^{14}$ cm$^{-2}$, $T\rm_{mb}$ approaches  a constant value when the temperature factor approaches 0 as shown in the small plot. }
\label{fig:oh_detect_paraspace}
\end{figure}

 We now consider the non-LTE cases.  As shown in Equation \ref{eq:65_col} and \ref{eq:67_col}, OH column density and  its uncertainty  are very sensitive to  $T\rm_{ex}$ through the function, $g(T_{\rm ex})=|T_{\rm ex}/(T_{\rm ex}-T_{ \rm bg})|$. It would be ten times lower for $g(T_{\rm ex})=1$ than that of $g(T_{\rm ex})=10$. But there exists a constraint on $g(T_{\rm ex})$ in order  that OH be detected with our sensitivity as shown in Figure \ref{fig:oh_detect_paraspace}. It requires a larger  deviation of $T_{\rm ex}/T_{\rm bg}$ from 1 for small $N$(OH) to be detected. Moreover, we are able to apply reasonable assumptions to different non-LTE cases. The following cases are clearly non-LTE when we consider the main lines (masers are ignored here). 

\begin{enumerate}[1)]

\item The existence of 1665 MHz line alone. 

\item The existence of 1667 MHz line alone.

\item  Both 1665 and 1667 MHz lines are present, but $R_{1667/1665}$ is out of the LTE range.

\end{enumerate} 

In case 1, the existence of the 1665 MHz line alone with the absence of the 1667 MHz  line implies the equality between excitation  and background temperature of 1667 line, $T_{\rm ex}(1667)=T\rm_{bg}(1667)$. Previous emission/absorption observations toward  continuum sources  revealed  $|T_{\rm ex}(1667)-$$T_{\rm ex}(1665)|$ $\sim 0.5 - 2$ K  \citep[e.g.,][]{crutcher79}.  We adopted $T_{\rm ex}(1665)-T_{\rm ex}(1667) = \pm$ 1.0 K, in which plus and minus are for emission and absorption of the 1665 MHz line, respectively. This adoption leads to  $|T_{\rm ex}(1665)/(T_{\rm ex}(1665)-T_{\rm bg}(1665))|$= 6.0 or 8.0 when $T\rm_{ex}(1667)=7.0$ K,  where we have adopted $T_{\rm bg}(1665)=T_{\rm bg}(1667)$  due to minor difference between them. 

 A similar strategy for calculations in case 2 was adopted. We cannot exclude the possibility of a detection limit  that leads to absence of 1665 MHz detection in case 2, since 1665 MHz line is generally weaker than 1667 MHz line. But the expected 1665 intensities ($T_A(1665)=5T_A(1667)/9$) are greater than 3 $\sigma$ rms  in 63\%  of case 2 clouds  and are greater than 2 $\sigma$ rms  in all clouds of case 2. Thus the assumption  of case 2 is reasonable. Uncertainties in case 1 and 2 are given with $|T_{\rm ex}(1667)-$$T_{\rm ex}(1665)|$  in the range of [0.5, 2.0] K. 

The value of $|T_{\rm ex}/(T_{\rm ex}-T_{\rm bg})|$ for case 1 and case 2 ranges from 4.3 to 11.5 with a median 7.04. We applied this median value  for all calculations in case 3. The uncertainty in case 3 is given with  $|T_{\rm ex}/(T_{\rm ex}-T_{\rm bg})|$ ranges of [4.3,11.5]. 

The optically thin assumption was applied to clouds under non-LTE conditions. This assumption is reasonable because there is no deviation from the `sum rule'  as presented in Section \ref{sec:sumrule}.  During the calculation of $N$(OH) of case 1,  equation \ref{eq:65_col} was employed.  Equation \ref{eq:67_col} was employed for cases 2 and 3. 

OH column densities of 117 clouds with main line emission were calculated. $N\rm(OH)$ ranges from $1.8\times10^{14}$ \cm2 to $1.1\times10^{16}$ \cm2 with a median  of $1.9\times10^{15}$ \cm2. Compared to OH column densities in clouds previously observed, this median value is about one order of magnitude larger than that determined explicitly through on/off observations toward 3C 133 and is more than 3 times the value in the W44 molecular cloud  \citep{myers75,crutcher79}. 

Two main uncertainties exist in the above assumptions of $p\rm_c$. The first originates from the distance ambiguity for directions toward the inner Galaxy.  For OH clouds associated with H{\sc i} self absorption, near distance is preferred  \citep{jackson02,roman09} as we have adopted. For other OH clouds, the distance ambiguity leads to a maximum difference of $p\rm_c$ between near and far distance of 0.57. Only 17 OH clouds are affected. The deviation factor of $N$(OH) caused by the distance ambiguity ranges from 0.049 to 2.0  with a median of 1.6.  

The second uncertainty is the difference between three-dimensional distribution of radio continuum emission over the entire Galaxy and the uniform distribution we assumed. \citet{beuermann85} reproduced  a three-dimensional model of the galactic radio emission from 408 MHz continuum map \citep{haslam82}, and found exponentially decreasing distribution of emissivities along galactic radius (4 kpc $<$ R $<$ 16 kpc) in the galactic plane. We adopted the detailed radial distribution in Fig. 6a of \citet{beuermann85}. The differences of $p\rm_c$ varies from -0.01 to 0.24 with a median value of 0.028. The deviation factor of $N\rm(OH)$ caused by three-dimensional model of radio emission ranges from $6\times 10^{-4}$ to 0.1 with a median value of 0.02. Thus the uncertainty from three-dimensional model is much smaller than that from intrinsic excitation temperature.

\subsection{H{\textsc{i}} Column Density}
\label{sec:hicolumn}

\hi  permeates in the Milky Way. The \hi spectrum includes all \hi contributions along a sightline and  toward the Galactic plane is  broad in velocity. It is difficult to distinguish a single \hi cloud without the help of special spectral features, e.g., \hi  narrow self absorption (HINSA) against a warmer \hi  background \citep[e.g.,][]{gibson00,li03}.

The excitation temperature (T$\rm_x$) and optical depth ($\tau$) of the
HINSA cloud is essential for deriving the \hi column density for a cloud with a
HINSA feature.   \citet{krco08} introduced a method of fitting the second derivative of
the \hi spectrum to derive the background spectrum and fitted
$\tau\rm_{HINSA}$ of HINSA cloud.  We combine the radiation
transfer equations in \citet{li03} and the analysis method in
\citet{krco08} for  calculation of N($\rm\hi$).  

 We assume a simple three-body
radiative transfer configuration with background warm \hi gas, cold \hi cloud, and 
foreground warm \hi gas.  The background \hi spectrum without absorption of cold \hi cloud,   
$T_{\mathrm{H{\textsc i}}}$, is related to the observed spectrum in which continuum has been removed,
$T_{\rm R}$ through the following equation (see details of equation 8 in  \citet{li03}), 

\begin{equation}
T_{\mathrm{H{\textsc i}}} = \frac{T_{\rm R} + (T_{\rm c}-T_{\rm k})(1-\tau_f)(1-e^{-\tau})}{1-p(1-e^{-\tau})}
\label{eq:hibackspec}
\end{equation}
where $T\rm_c$ represents the background continuum temperature contributed by the cosmic background and the Galactic continuum emission, $T_{\rm k}$ is the excitation temperature of the atomic hydrogen in the cold cloud, which is equal to the kinetic temperature, $\tau$ is the optical 
depth of the cold cloud. $\tau_f$ and $\tau_b$ are the optical depths of
warm \hi gas in front and behind the HINSA cloud. The total optical depth of warm \hi gas along the line of sight, 
$\tau_h=\tau_f+\tau_b$. $p$ is defined as the
fraction of background warm H{\sc i}, $p=\tau_b/\tau_h$. The value of $p$ is
calculated through
\begin{equation}
p=\int_{\rm behind}\Sigma(r)dr/\int_{\rm entire-LOS}\Sigma(r) dr, 
\label{eq:pfactor}
\end{equation}
where $\int_{\rm behind}\Sigma(r)dr$ and $\int_{\rm entire-LOS}\Sigma(r) dr$  are the integrated \hi surface densities behind the HINSA cloud and along the all line of sight. The surface density distribution in \citet{nakanishi03}  and the Milky Way rotation curve in \citet{brand93}  were used  for this calculation.   

\begin{figure}
  \includegraphics[width=0.48\textwidth]{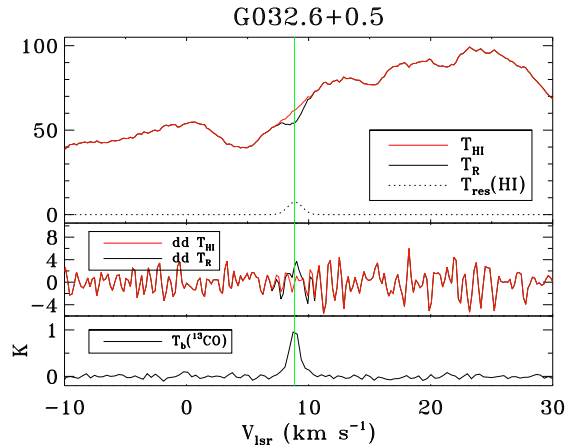}
  \caption{ $Top\ panel$: HINSA spectrum at $\sim 9$ km s$^{-1}$ along G032.6+0.5. The observed \hi spectrum ($T\rm_R$ ) and derived  \hi background spectrum ($T\rm_{\hi}$) are represented by the black and red solid lines, respectively. The dotted line shows the residual spectrum. During the fitting, the kinetic temperature of CO was fixed at 12 K. The original and fitted optical depths are 0.1 and  0.31.  $Middle\ panel$: Second derivatives of \hi and \hi background spectra. $Bottom\ panel$: Corresponding $^{13}$CO spectrum of HINSA. The green vertical line marks the fitted central velocity of the HINSA cloud from $^{13}$CO.}
\label{fig:G032.6+0.5_hinsa}
\end{figure}

We try to recover the background spectrum with Equation \ref{eq:hibackspec} to fit the second derivative as that in \citet{krco08}.  Information on the kinetic temperature is needed. HINSA features are pervasive in the Taurus molecular cloud. Analysis of pixels with both $^{12}$CO and $^{13}$CO emission in this region reveals a kinetic temperature in the range of [3,21] K, but concentrated in range of [6, 12] K. In most cases, we choose a fixed kinetic temperature of 12 K for $^{13}$CO that is widely used in molecular clouds  \citep{goldsmith08} and an initial HINSA optical depth of 0.1. The fitting result with a comparable thermal temperature of \hi gas to 12 K was chosen, otherwise we modify the initial parameter, e.g., relax the kinetic temperature in the range of [6,15] K as a free parameter.  An example is shown in Figure \ref{fig:G032.6+0.5_hinsa}.  The HINSA column density is  given by the fitted $\tau$ and FWHM of HINSA cloud, $\Delta V$ by  

\begin{equation}
N({\rm HINSA})=1.95\times10^{18}\tau\Delta V T_{\rm k}\    \rm cm^{-2},
\label{eq: colhinsa}
\end{equation} 
where $T\rm_k$ is the kinetic temperature of the HINSA cloud.  The HINSA column density  depends on the value of the kinetic temperature, thus uncertainties are given from kinetic temperature in the range of [6,15] K. 

HINSA traces the cold component of neutral hydrogen in a molecular cloud
 which may have a warm \hi halo \citep{andersson91}. We were able to determine \hi
column density of the \hi halo of molecular clouds with and without
HINSA features through the \hi spectra.  Due to the omnipresence of \hi
in the Galactic plane, we did not apply gaussian decomposition to \hi
profiles without HINSA feature. We derived the column density of \hi gas
through the integrated \hi intensity. The integrated \hi intensity of
recovered background spectrum was used for clouds with HINSA
features. With the assumption of low optical depth, the \hi column
density $N$(H{\sc i}) is given by
\begin{equation}
N({\rm H{\textsc i}})=1.82\times10^{18}\int T_{\rm b}d\upsilon\rm\  cm^{-2},  
\label{eq:hi_col}
\end{equation}
where the \hi intensity is obtained through integrating the velocity channels determined by OH lines. The effect of adopting different velocity widths  of \hi  is discussed in Section \ref{sec:oh_vs_hi}. $N$(H{\sc i}) derived using this method is limited by the optically thin assumption, the intensity contribution from clouds in neighboring velocities, and  \hi absorption features corresponding to OH emission lines  \citep[e.g.,][]{li03}. 

The HINSA column density, $N$(HINSA) derived in 52  clouds ranges from $8.4\times 10^{17}$ cm$^{-2}$ to  $4.0\times 10^{19}$ cm$^{-2}$ with a median value of $8.5\times 10^{18}$ cm$^{-2}$, which is  1/36 of the median \nhi\ of the \hi halos of these clouds. The median $N$(HINSA) is consistent with that derived in HINSA survey outside the Taurus Molecular Cloud Complex, log$_{10}(N\rm_{HINSA})$=18.8$\pm$ 0.53  \citep{krco10}. 

\subsection{CO Column Density}
\label{sec:cocolumn}

Six masks are defined for different detection cases. They are shown in Table \ref{table:detections}. When $^{12}\rm CO$,$^{13}\rm CO$ were detected simultaneously  (mask 3 and 6 in Table \ref{table:detections}), the clouds should be dense
molecular gas. In this case, $^{12}\rm CO$ is assumed to be optically
thick with $\tau_{12}\gg 1$. $T\rm_{ex}^{12}$, the excitation temperature of $^{12}\rm CO$, is given by

\begin{equation}
T_{\rm ex}^{12}=5.532\{{\rm ln}[1+\frac{5.532}{T_{\rm b}^{12}+0.819}]\}^{-1},
\label{eq:exciation_12co}
\end{equation}
 where $T\rm_b^{12}$ is the brightness temperature of $^{12}\rm CO$. 
 
The total column density of  $^{13}$CO, $N\rm_{tot}^{^{13}CO}$, is given by  \citep{qian12} \begin{equation}
N_{\rm tot}^{\rm {^{13}CO}}=3.70\times 10^{14} \int \frac{T_{\rm b}^{13}}{K} \frac{d\upsilon}{\rm km\ s^{-1}}  f_{u}f_{\rm \tau_{13}}f_{\rm b}\frac{1}{f_{\rm beam}} \  \rm cm^{-2},
\label{eq:col_13co}
\end{equation}
where $T\rm_b^{13}$ is the brightness temperature of $^{13}\rm CO$. $f_{\tau_{13}}=\int \tau_{13} d\upsilon/\int(1-e^{-\tau_{13}})d\upsilon$ is the correction factor of $\tau_{13}$, the optical depth of $\rm^{13}CO(1-0)$.  $\tau_{13}$ is given by 
\begin{equation}
\tau_{13}= -\rm ln(1-\frac{\textit{T}_b^{13}}{\textit{T}_b^{12}}) \ ,
\label{eq:tau_13co}
\end{equation}
in which $\tau_{12}\gg 1$ and $T\rm_{ex}^{12}=\textit{T}_{ex}^{13}$ are adopted. These are  reasonable when the excitation is dominated by collisions. The Galactic distribution of $\rm^{12}C/^{13}C$ ratio derived from synthesized observations of  CO,CN, and H$_2$CO in \citet{milam05}  was adopted to convert $N(\rm^{13}CO)$ to $N(\rm^{12}CO)$, $\rm^{12}C/^{13}C=6.21D_{GC}+18.71$, where $D\rm_{GC}$ is distance to the Galactic center. We adopted the Milky Way rotation curve of \citet{brand93}  to derive $D\rm_{GC}$.  Velocity dispersions and non-circular motions  \citep{clemens85}  are expected to affect the calculations of $D\rm_{GC}$  with an uncertainty of  $<$ 10\%, leading to an  uncertainty of $\sim$ 1\%  in $\rm^{12}C/^{13}C$.

For clouds with detection of only $^{12}\rm CO$ (masks 2 and 5 in Table
\ref{table:detections}), it is difficult to determine  $N(\rm CO)$
without observations of higher  lines, e.g., CO(2-1). With
the assumption that $^{12}\rm CO$ is optically thin, we derive a lower
limit. Adoption of a 3$\sigma$ detection of the $^{13}\rm CO$ will give
an upper limit to the column density.  Combining these two facts, we
adopted the average value of upper limit and lower limit for $N(^{12}\rm
CO)$. Under the optically thin assumption, the total column density of
$^{12}\rm CO$ can be expressed as

\begin{equation}
N_{\rm tot}^{\rm ^{12}CO}=3.57\times 10^{14} \int \frac{T_{\rm b}}{K} \frac{d\upsilon}{\rm km\ s^{-1}}  f_{u}f_{\tau}f_b\frac{1}{f_{\rm beam}} \  \rm cm^{-2},
\label{eq:col_12co}
\end{equation}
where $T\rm_b$ is the brightness temperature, $f_{u}=Q(T_{\rm
  ex})/g_{u}{\rm exp}(-h\nu/kT_{\rm ex})$ is the level correction
factor, $f_{\tau}=\int
\tau_{12}d\upsilon/\int(1-e^{-\tau_{12}})d\upsilon$ is the correction
factor of opacity and $f_{\tau}=1$ was adopted under optically thin
condition, $f_{\rm b}=1/[1-(e^{h\nu/kT_{\rm ex}}-1)/(e^{h\nu/kT_{\rm
      bg}}-1)]$ is the correction for the background, $f\rm_{beam}$ is
the beam filling factor of the cloud (assumed to be 1.0). 

A common excitation temperature of $T\rm_{ex}= 12$ K in molecular
regions (e.g., Taurus molecular cloud; Goldsmith et al. 2008) was
adopted during the calculation. $Q(T_{\rm ex})\approx T_{\rm ex}/2.76$ K
is the partition function. $g_{u}$ represents the degeneracy of the
upper transition level  and equals 3 for CO(1-0) transition. $\tau_{12}$
is the opacity of the $^{12}$CO(1-0) transition. $\tau_{12} \ll 1$ was
assumed, indicating an underestimation of $N(\rm^{12}CO)$ by a factor of
$\sim$ 5 when $\tau_{12} =5$.  $T\rm_{bg}$ is the background brightness
temperature, adopted to be 2.73 K.

A  $3\sigma$ upper limit on $\rm^{13}CO$ of 0.12 K  was adopted in  equations \ref{eq:col_13co} and \ref{eq:tau_13co} for calculation of the upper limit to $N(\rm^{12}CO)$. 

The column density of $^{12}$CO ranges from $2.7\times 10^{15}$ cm$^{-2}$ to $1.2\times 10^{18}$ cm$^{-2}$ with a median value of $6.5\times 10^{16}$ cm$^{-2}$. The median value of $N(\rm^{12}CO)$ for clouds with  both $^{12}$CO and $^{13}$CO emission is  $9.7\times 10^{16}$ \cm2, which is $\sim 11$ times  that in clouds with $^{12}$CO  emission alone.

\section{Results}
\label{sec:results}

\subsection{Sum Rule of Brightness Temperature }
\label{sec:sumrule}

\citet{robinson67} presented a brightness temperature `sum rule', which relates the intensities of the four OH ground-state transitions under the assumptions of small optical depths, a flat background continuum spectrum, and $T\rm_{ex}\gg 0.08$ K. The `sum rule' is,   
\begin{equation}
T_{\rm b}(1612)+T_{\rm b}(1720)=T_{\rm b}(1665)/5+T_{\rm b}(1667)/9.
\label{eq:tb_sum_rule}
\end{equation}

\begin{figure}
\centering
  \includegraphics[width=0.47\textwidth]{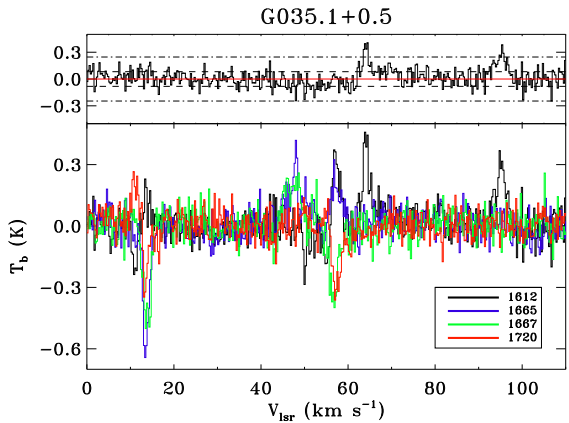}
  \caption{ Top panel:  residual spectrum representing  $T_{\rm b}(1612)+T_{\rm b}(1720)-T_{\rm b}(1665)/5-T_{\rm b}(1667)/9$. The 1 $\sigma$ and 3 $\sigma$ levels   of the spectrum are  indicated by in dashed and dash-dotted lines, respectively. Significant deviations are present for  the peaks around 64 km s$^{-1}$ and 90 km s$^{-1}$, which represent an evolved stellar maser associated with the infrared source  IRAS 18510+0203. Bottom panel:  OH  spectra for G035.1+0.5 are displayed as solid lines with different colors. }
\label{fig:g035.1+0.5_oh}
\end{figure}

Diffuse OH emission and absorption in the pilot region of SPLASH survey
followed this relation, (where ``diffuse" OH is defined as signal from
the extended molecular ISM, in which maser action is either absent or
very weak). Similar to \citet{dawson14}, we found no deviation
from the `sum rule' by more than 3 $\sigma$ for all diffuse OH emission and absorption. An example is shown in Figure
\ref{fig:g035.1+0.5_oh}. 

According to Appendix A,  the `sum rule' is valid for optically thin condition despite the existence of strong differences between the excitation temperatures of four OH lines. The deviation from the `sum rule' is dominated by the opacity of OH lines. No deviation from the `sum rule' confirms the validity of the optically thin assumption that has been used for calculation in Section \ref{sec:oh_col_den}. 

Maser amplification, which indicates strong non-LTE behavior and large
optical depth, leads to deviation from the `sum rule' as shown in Figure
\ref{fig:g035.1+0.5_oh}. Based on this fact, the `sum rule' can be used
as a filter for finding maser candidates.

\subsection{Comparison between Different Lines}
\label{sec:comp_dif_lines}

\hi is the tracer of atomic gas while CO is a tracer of molecular gas. C$^+$ emission traces both atomic and molecular gas. All OH clouds have associated \hi emission. Spectra of these lines toward G036.4+0.0 are shown in Figure \ref{fig:g036.4+0.0_spec}.  The statistics of clouds with C$^+$ and CO emission corresponding to OH  are listed in Table \ \ref{table:detections}. C$^+$ and  CO are present in 45\% and 80\% of all the OH clouds, respectively. We present a detailed comparison between column density  of H{\sc i}, CO line and intensity of \cp\ line with $N$(OH) in the following sections. 

\begin{deluxetable}{ccccccc}
\tablewidth{0pt}
\tablecaption{Summary of detections of all 151 OH clouds.  \label{table:detections}}
\tablehead{
Mask & OH  & C$^+$  & $^{12}$CO &  $^{13}$CO   & Number$^a$ & HINSA$^b$\\
 }

\startdata
  1      &     $\surd$    &     x     &  x  &  x  &17 & 1 \\
  2     &       $\surd$   &      x    &   $\surd$  &  x  &17  & 5\\
  3      &     $\surd$     &     x     &   $\surd$  &   $\surd$  & 50 & 24\\
   4    &     $\surd$     &     $\surd$      & x   &  x  &10 &  1\\
   5     &      $\surd$    &      $\surd$     &   $\surd$  &  x  & 9  & 2\\
   6     &      $\surd$    &    $\surd$       &    $\surd$ &    $\surd$ & 48 & 16
\enddata
\tablenotetext{a}{\ The number of clouds in each mask.}
\tablenotetext{b}{\ The number of HINSA detection in each mask.}
\end{deluxetable}

\subsubsection{Comparison between OH and \hi data}
\label{sec:oh_vs_hi}

There exist uncertainties in both the OH and H{\sc{i}} data. Thus we adopt  the IDL procedure $fitexy.pro$ for fitting, which considers uncertainties in both x and y directions (corresponding to $ \rm log N(OH)$ and $\rm log N$(H{\sc{i}}), respectively) during linear least-squares fitting.  The value of the fitted slope is larger than that when x error is not considered during fitting (see Table \ref{table:fittingpara} for comparison).

\begin{figure}
\begin{center}
  \includegraphics[width=0.48\textwidth]{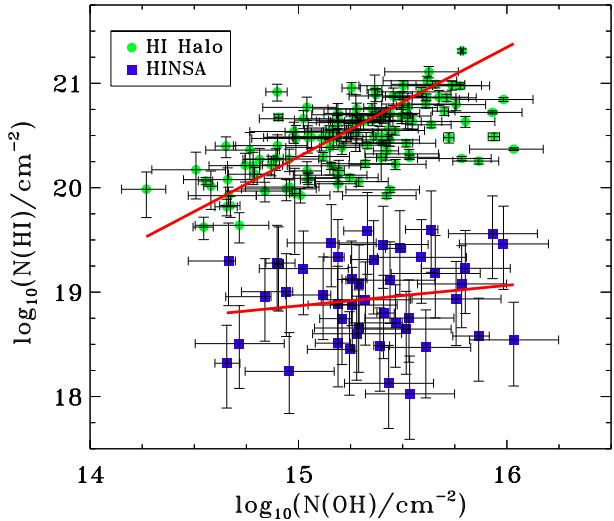}
  \caption{ H{\sc i} column density, $N(\rm \hi)$ in HINSA and \hi halo versus OH column density, $N$(OH).  Integrated emission spectra were used to derive $N(\rm \hi)$ in \hi halo. The two red solid lines show linear fit for two category of samples. }
\label{fig:hi_vs_oh}
\end{center}
\end{figure}

As shown in Table \ref{table:fittingpara},  the linear fit for clouds with HINSA in  Figure\ \ref{fig:hi_vs_oh}  is expressed as,  ${\rm log} N$(H{\sc{i}})=$0.20^{+0.20}_{-0.20}\ {\rm log} N(\rm OH)$+$15.9^{+3.1}_{-3.1}$. The value of the slope is 0.20 with an uncertainty of 0.20, indicating a weak correlation. The linear fit for H{\sc{i}} halo is ${\rm log} N$(H{\sc{i}})=$1.05^{+0.03}_{-0.03}\ {\rm log} N(\rm OH)+4.57^{+0.43}_{-0.43}$. The value of the slope is 1.0 with an uncertainty of 0.028, indicating a strong correlation.  These results show that the correlation between OH and warm \hi is better than that between OH and HINSA. This seems to conflict with the fact that cold \hi rather than warm \hi is mixed with molecular gas  \citep[e.g.,][]{goldsmith05}. The explanation may be in part the following. 

\begin{enumerate}[1)]
\item \citet{goldsmith05} studied local dark clouds not in directions toward the Galactic plane.  There was thus little velocity ambiguity for these observations.  In the present study of sightlines along the Galactic plane, the OH velocity width was used for calculating $N$(H{\sc{i}}) for  \hi halo gas. This may produce a bias toward apparent  correlation. 

We note that the velocity width for the \hi halo is always larger than that of molecular tracers, e.g., CO. But no strong correlation between them is found \citep{andersson91}. \citet{lee12}  compared correlation between derived  $N$(H{\sc i}) with different \hi widths and 2MASS extinction, finding the best correlation between \hi emission and extinction at 20 \kms around the CO velocities.   The width is much larger than linewidth found for CO, OH, and \hi self-absorption, making  such a correlation suspicious. The logic here almost runs in a circle if one tries to study the behavior of \hi associated \h2 by only looking at the velocity range best associated with \h2.
The analysis toward clouds in the Galactic plane is more complicated. Firstly, the extinction along a sightline represents sum of all clouds in this sightline.  Secondly, the brightness temperature in velocity range of a molecular cloud  would be diluted by extended emission of other clouds in this sightline. Thus, we adopted the OH velocity range to calculate  \hi column density of to examine the possible \hi halo around our targets. 

\item Two correlations with contrary behavior  exist between HINSA and OH. Firstly, HINSA content has a positive correlation with increasing molecular  cloud size, which can be represented by N(OH). Secondly, \hi is depleted to form \h2, leading decreasing  N(HINSA) as the proportion of OH increases. If these two factors are comparable, the absence of correlation between HINSA column density and OH column density is expected. 

\end{enumerate}

A feature in Figure\ \ref{fig:hi_vs_oh} is that $N$(H{\sc{i}}) trends to saturate at $1.0 \times 10^{21}$ cm$^{-2}$ when $N\rm(OH)) \gtrsim 4.5  \times 10^{15}$ cm$^{-2}$. A similar feature is seen in Spider and Ursa Major cirrus clouds, where the asymptotic value of $N$(H{\sc{i}}) is $5\times 10^{20}$ \cm2 when $N(\rm OH)> 0.25\times 10^{14}$ \cm2 \citep{barriault10}. The asymptotic values of  $N$(H{\sc{i}}) between this study and \citet{barriault10} are consistent but the critical value of $N(\rm OH)$ in this paper is larger by two orders of magnitude. This asymptotic behavior implies that the mass of \hi halo will be same for different clouds when the molecular core is large enough.  This behavior also implies that a portion of molecular gas may be not well traced by  \hi in the halo. 

\begin{deluxetable*}{llllll}
\tablewidth{0pt}
\tablecaption{ Fitting information in Figure \ref{fig:hi_vs_oh} (ID 1 and 2) and Figure \ref{fig:cp_vs_oh} (ID 3). \label{table:fittingpara}}
\tablehead{
 ID & Parameters  & Category & Fitted slope$^a$  & Fitted intercept$^a$  & Fitted slope (SLLS)$^b$
 }
\startdata
 1 & N($\rm \hi$) vs N(OH) & \hi halo      &  1.05 $\pm$ 0.03 &  4.57 $\pm$ 0.43  & 0.58$\pm$ 0.07   \\
 2 & N($\rm \hi$) vs N(OH) & HINSA    &  0.20 $\pm$ 0.20 & 15.9 $\pm$ 3.1 & 0.14 $\pm$ 0.18 \\   
 3 &  I(C$^+$) vs N(OH)  & CO-dark &  0.63 $\pm$ 0.46  & -9.44 $\pm$ 7.12 & 0.51 $\pm$ 0.20
 \enddata
 
 \tablenotetext{a} { Fitted slope and intercept considering uncertainties in both X and Y coordinates.}
\tablenotetext{b} { Fitted slope with standard linear least-square (SLLS) method when X error is not considered.}
\end{deluxetable*}

\subsubsection{Comparison between OH and CO data}
\label{sec:oh_vs_co} 

\begin{figure}
\centering
  \includegraphics[width=0.47\textwidth]{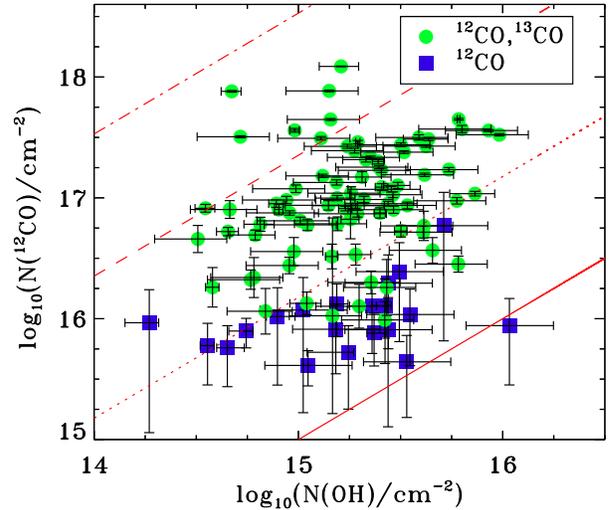}
  \caption{ Comparison of $N(\rm^{12}CO)$  with $N(\rm OH)$ on a log-log scale for 100 clouds. The green filled circles represent 24 OH clouds with $^{12}$CO emission alone. The blue filled squares represent 76 OH clouds with both $^{12}$CO and $^{13}$CO emission. Solid, dotted, dashed, and dash dotted lines represent $N$(CO)/$N$(OH) values of 1, 15, 225, and 3375, respectively. The error bars of $N(^{12}\rm CO)$ for clouds with $^{12}\rm CO$ detections represent upper and lower limits. For clouds with detections of both $^{12}\rm CO$ and $^{13}\rm CO$,  statistical uncertainties of the CO spectrum are given. The uncertainty of $N\rm(OH)$ is the same as that described in Figure \ref{fig:hi_vs_oh}.} 
\label{fig:co_vs_oh}
\end{figure}

The OH column densities are compared with $^{12}$CO column densities in Figure\ \ref{fig:co_vs_oh}. There is no obvious correlation between $N$($^{12}$CO)  and  $N$(OH).  A possible reason for this is that OH may reveal larger fraction of molecular gas than CO,  e.g., the ``CO-dark" gas component, the fraction of which can reach 0.3 even in CO emission clouds \citep{wolfire10}.   

The  ratio between CO and OH column density, $N$(CO)/$N$(OH) varies from 3.7 to $1.6\times 10^3$ with a median value of 59  for clouds with both $^{12}$CO and $^{13}$CO emission. It varies from 0.81 to 52 with a median value of 7.1 for clouds with only $^{12}$CO emission.  These results confirm that OH will be depleted to form CO, resulting in larger $N$(CO)/$N$(OH) ratios in more massive molecular clouds.

\subsubsection{Comparison between OH and C$^+$ Emission}
\label{sec:cp_vs_oh}

The C$^+$ 158 $\mu$m fine-structure transition is sensitive to column density, volume density, and kinetic temperature of \hi and \h2, making it difficult to determine \cp\ column density based on present data. We adopted C$^+$  intensity rather than C$^+$ column density as a parameter for comparison. C$^+$ emission can be produced by both photon-dominated regions (PDRs) and the ionized gas in H {\sc{ii}} regions.  The average  ratio of C$^+$ emission from H {\sc{ii}} and PDRs in IC 342 is 70:30 \citep{rollig16}. This fact is included in estimating  the uncertainty of  C$^+$ intensity. We compared the relation between I(C$^+$) and N(OH) in CO-dark clouds (mask 4) and molecular clouds (mask 5 and 6).

\begin{figure*}
\centering
  \includegraphics[width=0.47\textwidth]{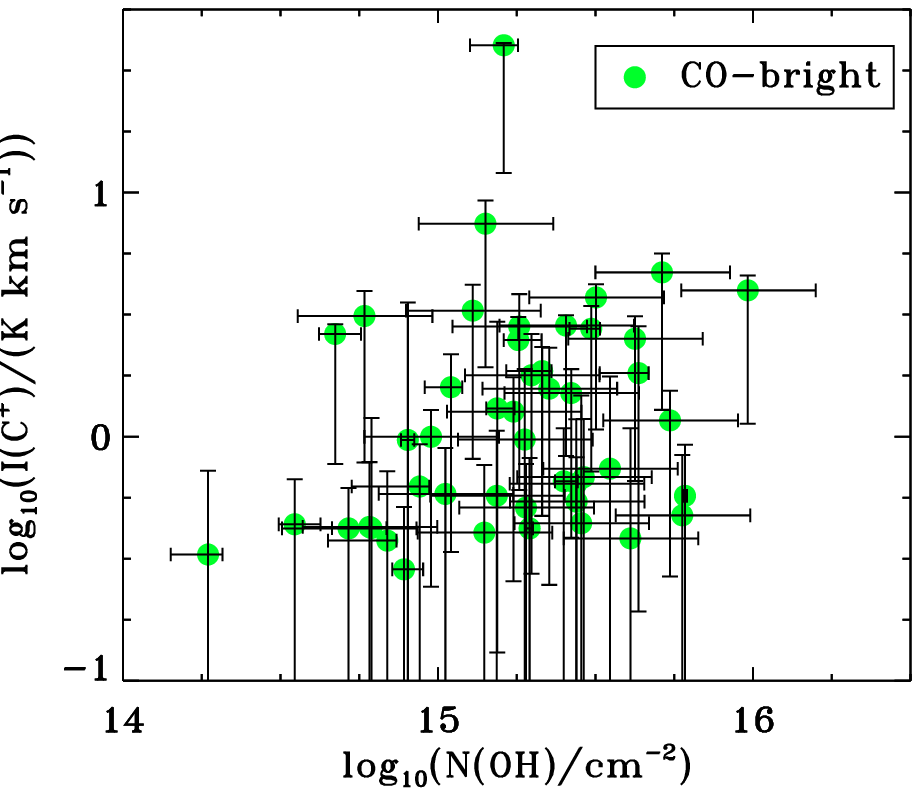}
  \includegraphics[width=0.47\textwidth]{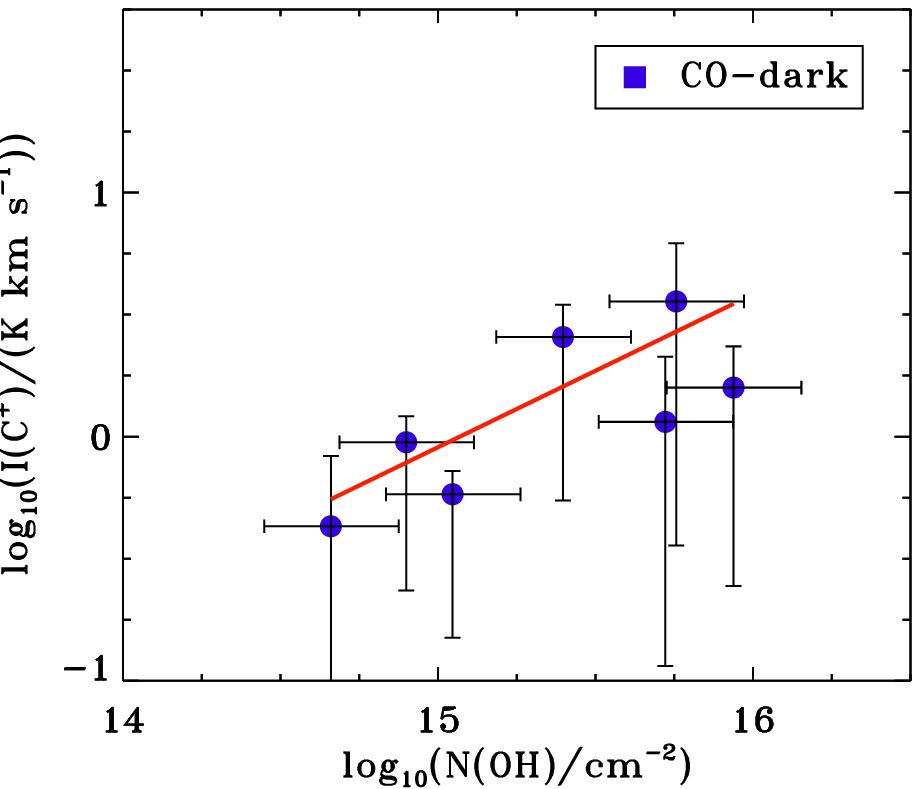}
  \caption{ Comparison of \cp\ intensity  with OH column densities on a log-log scale for 54 clouds.  Statistical uncertainty of C$^+$ spectrum is shown in error bar of C$^+$ intensity. The uncertainty in $N\rm(OH)$ is the same as that described in Figure \ref{fig:hi_vs_oh}. $Left:$ Forty-seven CO-bright clouds with CO emission are indicated by green filled circles.  $Right:$ Seven CO-dark clouds without CO emission are indicated by  blue circles. The red solid line represents a linear fit to these clouds. }
\label{fig:cp_vs_oh}
\end{figure*}

The clouds were divided into two categories, CO-bright and CO-dark. As seen in Figure\ \ref{fig:cp_vs_oh}, no correlation was found between $I(\rm C^+)$ and $N$(OH) for CO-bright category.  But this comparison is limited by the large uncertainty  in the C$^+$ data and the fact that C$^+$ traces both atomic and molecular components.    For the CO-dark category, the fitted slope is $0.63\pm 0.46$ (Table \ref{table:fittingpara}) with fitted Chi-squre  $\chi^2=1.84$, indicating a linear correlation. Based on the fact that \cp\ is a good tracer of \h2 in not well-shielded gas \citep[e.g.,][]{pineda13}, the correlation is consistent with the suggestion that OH is a better trace of \h2 than CO in diffuse clouds, though the sample size is small.  

\section{Discussion}
\label{sec:discussion}

\subsection{OH Column Density}
\label{sec:oh_column_density_dis}

\citet{crutcher79} found that OH column density is proportional to the
extinction following $N({\rm OH})/A_{\rm V} \approx 8\times 10^{13}$
cm$^{-2}$ mag$^{-1}$ in $A\rm_V$ within the range of 0.4--7 mag, which
implies OH/H $\approx 4\times 10^{-8}$. The minimum value of $N$(OH)
found in this study is $1.8\times 10^{14}$ cm$^{-2}$, which
corresponds to an extinction of 2.3 mag. If the $N({\rm OH})/A_{\rm
  V}$ relation extends to higher extinction, the maximum and median
$N$(OH) values would correspond to 138 mag and 24 mag.  The value of 24
mag is comparable to the largest extinction in Taurus cloud \citep{pineda10} 
while the value of 138 mag requires more dense gas. One possible
reason is that the value of $N({\rm OH})/A_{\rm V}$ is larger than
$8\times 10^{13}$ cm$^{-2}$ mag$^{-1}$ when $A\rm_V > 7$ mag.  The other reason
is that we may have overestimated $N$(OH).

Some satellite lines of OH show `flip' feature inverting from emission
to absorption at a velocity. An example is shown in
Figure\ \ref{fig:sate_anomaly}. This feature can be interpreted with
overlap of infrared transition of OH and implies a transition column
density of $N\rm_{OH}/\Delta V \approx 10^{15}$ cm$^{-2}$ km$^{-1}$ s,
where $\rm \Delta V$ is the full width at half-maximum of OH line \citep[e.g.,][]{crutcher77,brooks01}. These `flip' features were
found in three clouds of this survey. The values of N(OH) for `flip'
feaure in these clouds can be derived.  To compare the results that
derived with non-LTE assumption in Section \ref{sec:oh_col_den}, we
listed calculated $N$(OH) with two different methods in Table
\ref{table:OHcolden_discuss}. The results are consistent within a
factor of 2.5, confirming the validity of our calculation of $N$(OH) in
Section \ref{sec:oh_col_den}.

\begin{deluxetable}{lllll}
\tablewidth{0pt}
\tablecaption{ OH column density for three clouds with `flip' of satellite lines.\label{table:OHcolden_discuss}}
\tablehead{
Sightline   &  V$\rm_{lsr}$   &  $\Delta\rm V_{1667}$ &  N(OH)$\rm_{non-LTE}$$^a$     &  N(OH)$\rm_{flip}$$^b$     \\
         &   km s$^{-1}$     &     km s$^{-1}$     &    $10^{15}$ cm$^{-2}$    &    $10^{15}$ cm$^{-2}$    
         }
\startdata
G033.8-0.5    &  11.0  &  1.3  &   $0.56_{-0.1}^{+0.04}$  &   1.3  \\
G033.8+0.0   &   55.2  & 2.7  &   $6.3^{+4.1}_{-2.4}$   &   2.7  \\
G035.1+1.0   &   13.2  & 1.6  &   $1.3^{+0.1}_{-0.3}$    &  1.6  
\enddata

\tablenotetext{a} { N(OH) calculated with non-LTE method in Section \ref{sec:oh_col_den}.}
\tablenotetext{b} { N(OH) calculated with $N\rm_{OH} \approx 10^{15}*\Delta V_{1667}$ cm$^{-2}$.}
\end{deluxetable}

\subsection{CO-dark Molecular Gas and Atomic/Molecular Transition }
\label{sec:lines_trace_dmg}

A large fraction of molecular gas is expected to exist in the transition
region between the fully molecular CO region and the purely atomic \hi
region. The molecular gas in this region, which is called the ``CO-dark
molecular gas" (DMG), cannot be traced by CO, but is associated with
ions or molecules that are precursors of CO formation.  C$^+$ and OH are
two of them. One example of a DMG cloud is shown around a velocity of
41.9 \kms\ in Figure \ref{fig:g036.4+0.0_spec}. There exist OH and C$^+$
emission without corresponding $^{12}$CO  detections with a sensitivity of
0.07 K. Twenty seven DMG clouds, which comprise 18\% of all OH
clouds, are identified as shown in Table \ref{table:detections}.  This
fraction is smaller than that of $\sim 0.5$ found in pilot OH survey
toward Outer Galaxy  \citep{allen15}. The CO sensitivity of 0.05 K in
\citet{allen15} is comparable to that in this study. But the OH
sensitivity of $\sim 3.0$ mK in \citet{allen15} is ten times lower than
that in this study. This comparison indicates that the DMG fraction detected
depends on OH sensitivity. Higher DMG fraction is expected with higher
OH sensitivity. 

The DMG can  be a significant fraction  even in clouds with CO emission  \citep{wolfire10}  and in observations \citep[e.g.,][]{grenier05,langer14,tang16}. The fact that the DMG  can be traced by OH may explain the absence of correlation between OH and CO in Figure \ref{fig:co_vs_oh}.

\cp\ is the main reservoir of carbon in diffuse gas. It converts to
CO quickly through \cp-OH chemical reactions once OH is formed \citep[e.g.,][]{van88}.
 Thus \cp\ and OH are expected to have a
tight correlation. As shown in Figure \ref{fig:cp_vs_oh}, a $I(\rm
C^+)$-$N$(OH) correlation may exist for DMG clouds but not for all
clouds.

The atomic to molecular transition occurs in the DMG region. Though
HINSA other than warm \hi gas is associated with molecular formation,
the \hi halo outside the DMG region provides shielding from UV
radiation. The asymptotic value of \nhi, $1.0\times 10^{21}$ \cm2
in Figure \ref{fig:hi_vs_oh}, corresponding visual extinction
$A\rm_V$ of 0.5 mag,  approaching extinction $A\rm_V = 0.5-1$ mag
\citep[e.g.,][]{van88} required to provide effective shielding for
CO formation. This extinction of 0.5 mag is much larger than that
required for forming abundant \h2, which has a large self-shielding
coefficient and can be the dominant form of hydrogen even when
$A\rm_V>0.02$ mag \citep{wolfire10}. Thus a large fraction of DMG will
exist before abundant CO formation.

\section{Conclusions}
\label{sec:conclusion}

We have obtained OH spectra of four 18 cm lines toward 51 GOT C+ sightlines with the Arecibo telescope. Using Gaussian decomposition, we identified 151 OH components. A combined analysis of OH, CO,\hi, and HINSA reveals the following results. 

\begin{enumerate}[1)]
\item   OH emission is detected in both main and satellite lines in the inner Galactic plane but is only detected in the main lines in the outer galaxy. A large fraction of detected main lines show absorption features in the inner galaxy but no OH absorption feature was found in the outer galaxy. This is in agreement with more  molecular gas and a higher level of continuum background emission being present in the inner galaxy than  in the outer galaxy. 

\item   There is no deviation from the  `sum rule'  by more than 3$\sigma$ for all of the detected diffuse OH emission, suggesting small opacities of OH lines for clouds in the Galactic plane.

\item The \hi column density \nhi\ in the \hi cloud halos  has an obvious correlation with the OH column density $N$(OH) following ${\rm log} N$(H{\sc{i}})=$1.05^{+0.03}_{-0.03}\ {\rm log} N(\rm OH)+4.57^{+0.43}_{-0.43}$. \nhi\ reaches an asymptotic value of $1.0\times 10^{21}$ \cm2 when $N\rm (OH) > 4.5\times 10^{15}$ \cm2. 

\item No correlation was found between the cold \hi column density $N$(HINSA)  from \hi narrow self-absorption feature and $N$(OH).

\item  $N\rm(OH)$/$N\rm(CO)$ ratios are ten times lower in translucent clouds with only $^{12}$CO detection than  in dense clouds with both $^{12}$CO and $^{13}$CO detections. This confirms that OH is depleted to form CO. 
No correlation between $N\rm(OH)$ and $N\rm(CO)$ was found. 

\item   A  weak correlation was found between C$^+$ intensity $I\rm(C^+)$ and $N$(OH) for CO-dark molecular clouds. This is consistent with OH  being better tracer of  \h2 in diffuse molecular clouds as \cp\ traces \h2 well in not well-shielded gas.  No correlation was found for $I\rm(C^+)$ and $N$(OH) for CO-bright molecular clouds.

\end{enumerate}

\section*{Acknowledgments}

We thank anonymous referee for significantly improving this paper by pointing out  uncertainties that had been missed.  This work is supported by  International Partnership Program of Chinese Academy of Sciences  No.114A11KYSB20160008,  the Strategic Priority Research Program ``The Emergence of  Cosmological Structures" of the Chinese Academy of Sciences, Grant No.  XDB09000000, National Natural Science Foundation of China No. 11373038,  and National Key Basic Research Program of China ( 973 Program )  2015CB857100, the China Ministry of Science.  This work was carried out in part at the Jet Propulsion Laboratory, which is operated for NASA by the California Institute of Technology. M. K. acknowledges the support of Special Funding for Advanced Users, budgeted and administrated by Center for Astronomical Mega-Science, Chinese Academy of Sciences (CAMS). N.\ M.\ M.-G.\ acknowledges the support of the Australian Research Council through grant FT150100024. The Arecibo Observatory is operated by SRI International under a cooperative agreement with the National Science Foundation (AST-1100968), and in alliance with Ana G. M\'{e}ndez-Universidad Metropolitana, and the Universities Space Research Association.  CO data were observed with the Delingha 13.7m telescope of the Qinghai Station of Purple Mountain Observatory. We appreciate all the staff members of the Delingha observatory and Zhichen Pan for their help during the observations.

\appendix

\section{Derivation of Sum Rule }
\label{sec:appendix}
\begin{figure}
  \centering
  \includegraphics[width=0.47\textwidth]{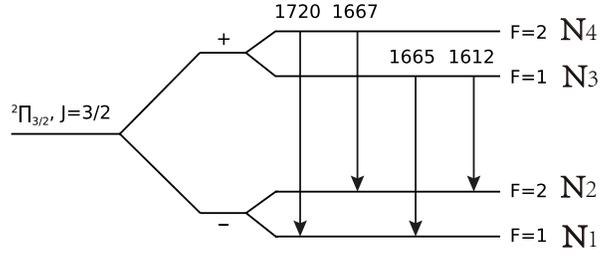}
\caption{ Energy levels responsible for the OH lines. This figure is reproduced from \citet{dawson14}. }
\label{fig:rp-pth}
\end{figure}

The column  densities of upper and  lower levels of each transition, $N\rm_i$ and $N\rm_j$, are  related by 

\begin{equation}
\frac{N_{\rm i}}{N_{\rm j}}=\frac{g_{\rm i}}{g_{\rm j}}e^{-h\nu/kT_{\rm ij}}
\label{eq:column_1}
\end{equation}
where $T\rm_{ij}^{ex}$ is the excitation temperature of OH transition line, $N\rm_i$ and $N\rm_j$ are partition weight the OH transition, and $g\rm_i$ and $g\rm_j$ are the statistical weights of i and j levels, respectively. For OH lines, $g=2F+1$, where F is    the total angular momentum quantum number.  The transition of 1612 MHz gives 
\begin{equation}
\frac{N_3}{N_2}=\frac{g(F=1)}{g(F=2)}e^{-h\nu_{1612}/kT_{1612}}.
\label{eq:column_2}
\end{equation}
$N_3$ and $N_2$ are upper and lower level of 1612 MHz line as shown in Figure \ref{fig:rp-pth}.

Similarly, $\frac{N_3}{N_1}$,$\frac{N_4}{N_2}$, and $\frac{N_4}{N_1}$ are derived from 1665, 1667, and 1720 MHz transitions, respectively. Considering the fact that $\frac{N_3}{N_2}\times\frac{N_4}{N_1}$=$\frac{N_3}{N_1}\times\frac{N_4}{N_2}$, we have 

\begin{equation}
\frac{\nu_{1612}}{T_{1612}}+\frac{\nu_{1720}}{T_{1720}}=\frac{\nu_{1665}}{T_{1665}}+\frac{\nu_{1667}}{T_{1667}}.
\label{eq:nu_tex}
\end{equation}

The optical depth $\tau_{\nu}$ at frequency $\nu$ is  given by
\begin{equation}
\tau_{\nu}=\frac{c^2}{8\pi}\frac{A_{\rm ij}}{\nu^2}N_{\rm i}(e^{h\nu_{\rm ij}/kT_{\rm ij}}-1)\phi(\nu). 
\label{eq:op_depth}
\end{equation}

With  $h\nu_{\rm ij}/kT_{\rm ij}\ll 1$ for OH lines in general, we derive  
\begin{equation}
\frac{\nu_{\rm ij}}{T_{\rm ij}}=\frac{8\pi}{c^2}\frac{k}{h}\frac{\nu_{\rm ij}^2}{A_{\rm ij}N_{\rm i}}\int \tau_\nu d\nu.
\label{eq:nu_tij}
\end{equation}
When most OH molecules are in the ground state of $^2\Pi_{3/2}(J=3/2)$, the total OH column density, $N\rm_{OH}$, is  sum of  molecules in four energy levels, $N\rm_{OH}$=$N_1+N_2+N_3+N_4$=$16/3N_3$=$16/5N_4$. $\int \tau_{\nu}d\nu =\tau_{\nu}^{\rm peak}\Delta V/0.93$, where $\tau_{\nu}^{\rm peak}$ is peak optical depth and $\Delta V$ is full width at half maximum of transition line. Combining  equations \ref{eq:nu_tex}, equation \ref{eq:nu_tij}, and  values of four OH transition coefficients, we derive

\begin{equation}
\tau_{1612}+\tau_{1720}=\tau_{1665}/5+\tau_{1667}/9,
\label{eq:sum_rule}
\end{equation}
where $\tau_{1612}$,$\tau_{1720}$,$\tau_{1665}$,and $\tau_{1667}$ are peak optical depth of four OH transitions. The only requirement for Equation \ref{eq:sum_rule} is $h\nu_{\rm ij}/kT_{\rm ij}\ll 1$. Thus Equation \ref{eq:sum_rule}  is valid even for non-LTE conditions. 

The brightness of emission line, $T\rm_b$, is calculated through $T_{\rm b}=(T_{\rm ex}-T_{\rm bg})(1-e^{-\tau})$, where  $T\rm_{bg}$ is background continuum temperature, $\tau$ is optical depth of transition line. When  $T_{\rm ex}-T_{\rm bg}$  is the same for the four OH lines, which is  valid under LTE conditions, we derive the `sum rule' for brightness temperature under optically thin conditions

\begin{equation}
T_{\rm b}(1612)+T_{\rm b}(1720)=T_{\rm b}(1665)/5+T_{\rm b}(1667)/9.
\label{eq:sum_rule2}
\end{equation}
 
\begin{figure}
  \centering
  \includegraphics[width=0.47\textwidth]{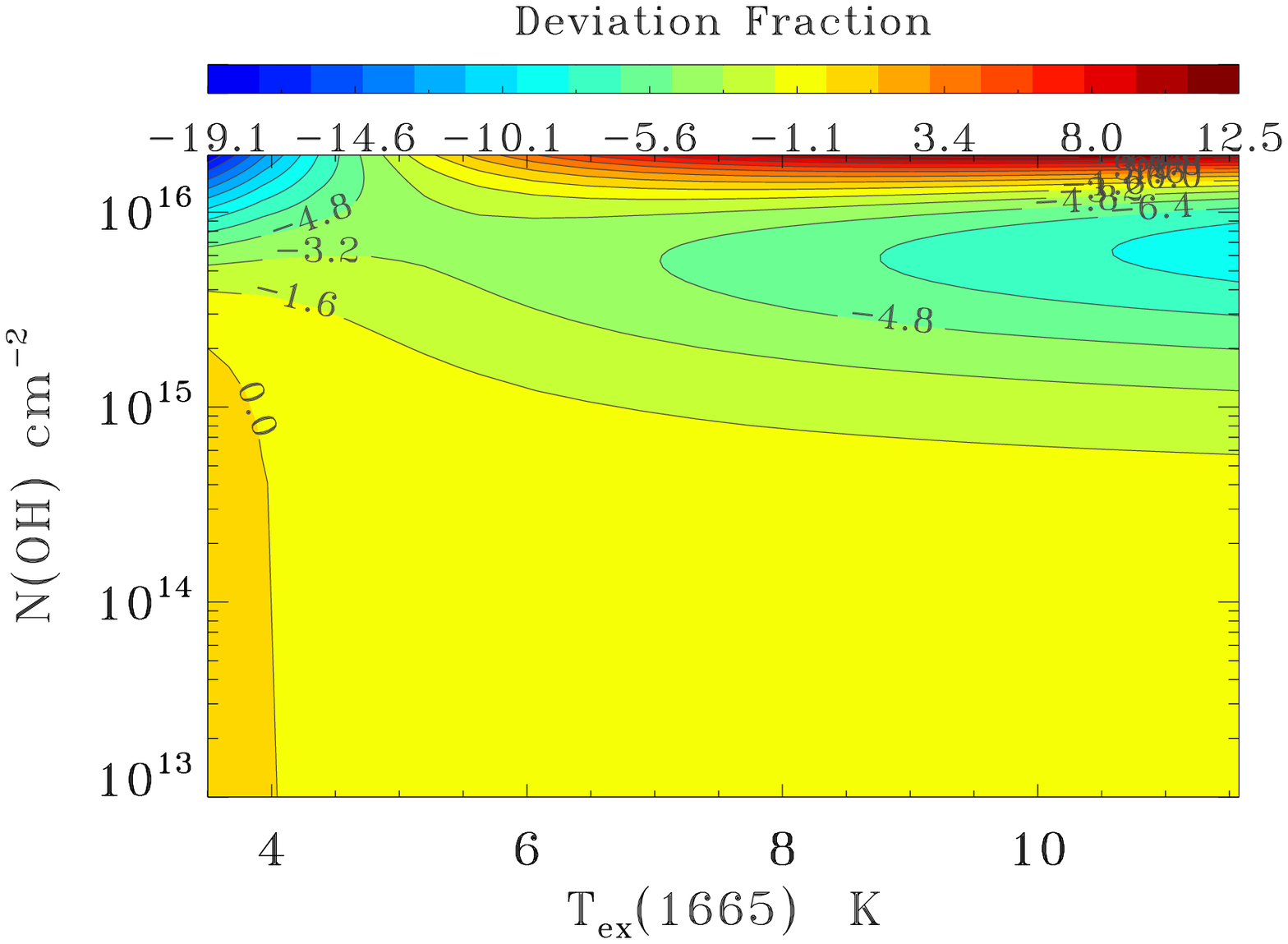}
  \includegraphics[width=0.47\textwidth]{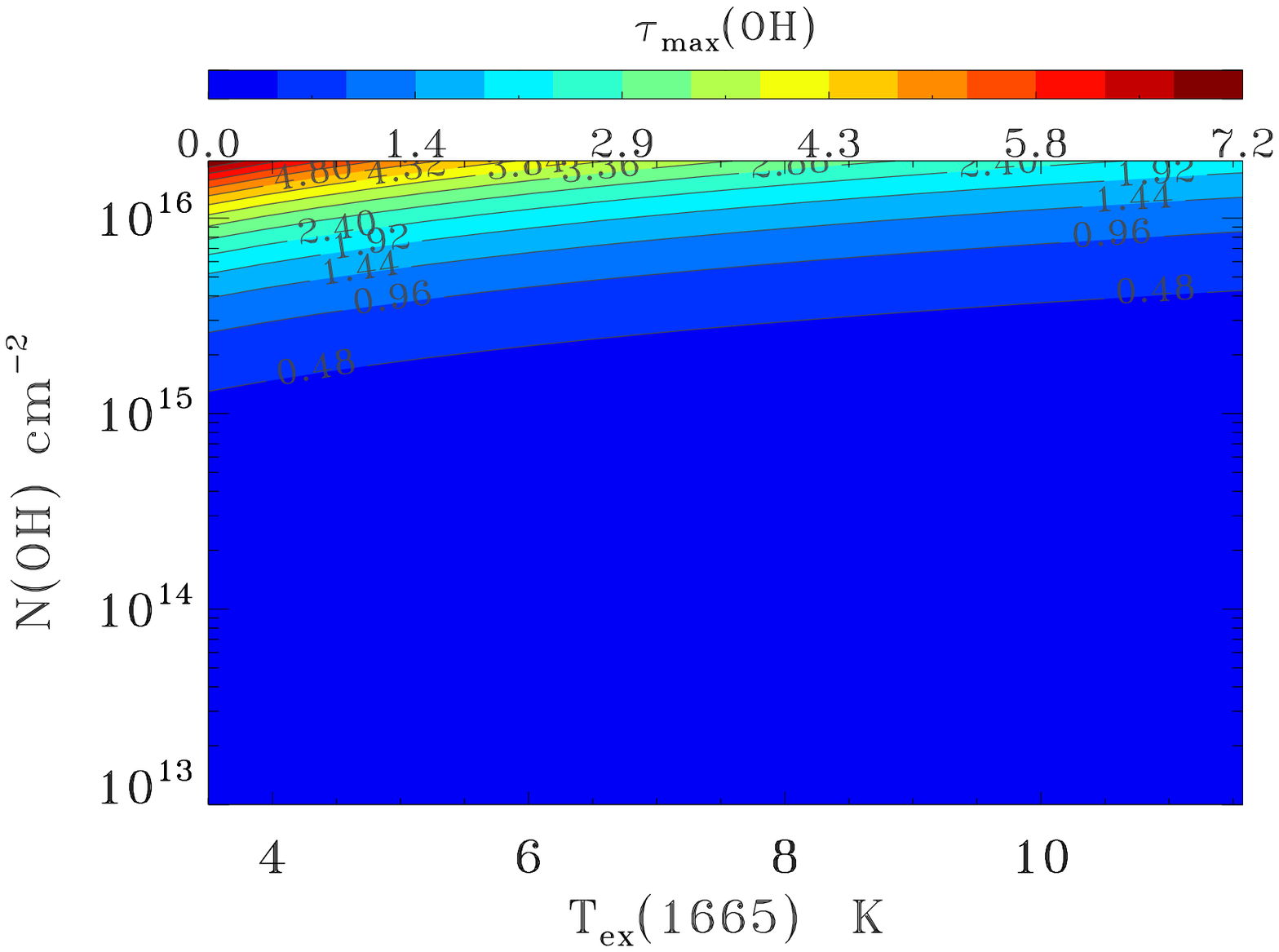}
\caption{ \textit{Top:}   Contour of deviation fraction of  the brightness sum rule,  {$f$=[$T\rm_b(1612)$+$T\rm_b(1720)$-$T\rm_b(1665)/5$-$T\rm_b(1667)/9$]/$\sigma\rm_{spec}$}, as a function of $T\rm_{ex}(1665)$ and total OH column density $N\rm_{tot}$(OH).   $\sigma\rm_{spec}=42$ mK is the rms of the summed spectra. $T_{\rm ex}(1667)-T_{\rm ex}(1665)=1.0$ K,  $T\rm_{ex}(1720)=4.0$ K, and continuum brightness temperature at 1667 MHz of 5.0 K are assumed. $T\rm_c$ at 1612 MHz is calculated through $T\rm_c(1612)$=3.1+$(T\rm_c(1667)-3.1)(1612/1667)^{2.1}$.  The continuum brightness at 1720 MHz is calculated using similar method.  \textit{Bottom:} Contour of the maximum optical depth of OH lines. 
 }
\label{fig:sum_rule_contour}
\end{figure}

To estimate the  contribution of non-LTE and optical depth leading to deviation from Equation \ref{eq:sum_rule2}, we plot deviation fraction and maximum optical depth of OH as function of $T\rm_{ex}(1665)$ and $N$(OH) in Figure \ref{fig:sum_rule_contour}. The deviation from the sum rule (DSR) is significant when $N$(OH) $\ge 2.0\times 10^{15}$ \cm2. In this parameter space  with significant deviation, the  optical depth of OH lines $\tau\rm_{max}$(OH) is greater than 0.5. We conclude that the condition of large optical depth results in significant DSR. 
  
\end{document}